\documentclass[11pt]{article}
\usepackage{amsmath}
\usepackage{amssymb}
\usepackage{babel}
\usepackage{latexsym}
\usepackage{times}
\usepackage{mathptm}
\begin{document}
\protect\newtheorem{principle}{Principle}[section]
\protect\newtheorem{theo}[principle]{Theorem}
\protect\newtheorem{prop}[principle]{Proposition}
\protect\newtheorem{lem}[principle]{Lemma}
\protect\newtheorem{co}[principle]{Corollary}
\protect\newtheorem{de}[principle]{Definition}
\newtheorem{fo}[principle]{Consequence}
\newtheorem{rle}{Rule}
\newtheorem{rem}[principle]{Remark}
\newtheorem{ex}[principle]{Example}
\newcommand{\HH} {{\frak H}}
\newcommand{\df} {decoherence functional }
\newcommand{\dfs} {decoherence functionals }
\vspace*{1.6cm}
\noindent {\Large The representation theory of
decoherence functionals in history quantum theories} \\ \\ \\ \\
{\large Oliver Rudolph}  \\ \vskip.0001in
\noindent {\normalsize Theoretical Physics Group, Blackett Laboratory,}
{\normalsize Imperial College of Science, Technology and
Med\-i\-cine,}
{\normalsize Prince Consort Road,}
{\normalsize London SW7 2BZ, United Kingdom} \\ \\
{\normalsize Talk given at the fourth IQSA biannual meeting
\emph{Quantum Structures '98}, August 30 - September 5, 1998,
Liptovsk\'y J\'an, Slovakia}
\\ \\
\\ \\
\normalsize
\noindent \emph{Abstract} In the first part of this paper
the general perspective of history quantum theories is reviewed.
History quantum theories provide a conceptual
and mathematical framework for formulating quantum theories
without a globally defined Hamiltonian time evolution and for
introducing the concept of space time event into quantum theory.
On a mathematical level a history quantum theory is characterized
by the space of histories, which represent the space time events,
and by the space of decoherence functionals which represent the
quantum mechanical states in the history approach. \\
The second part of this paper is devoted to the study of the
structure of the space of decoherence functionals for some
physically reasonable spaces of histories in some detail. The
temporal reformulation of standard Hamiltonian quantum theories
suggests to consider the case that the space of histories is given
by (i) the lattice of projection operators on some Hilbert space or --
slightly more general -- (ii) the set of projection operators in some
von Neumann algebra. In the case (i) the conditions are
identified under which decoherence functionals can be represented
by, respectively, trace class operators, bounded operators or
families of trace class operators on the tensor product of the
underlying Hilbert space by itself. Moreover we shall discuss the
naturally arising representations of decoherence functionals as
sesquilinear forms. The paper ends with a discussion of the
consequences of the results for the general axiomatic framework of
history theories.
\\
\section{Introduction}

\indent In standard textbook Hamiltonian
quantum mechanics the time variable is
fixed from the outset as the variable conjugate to the
Hamiltonian. A quantum mechanical
system is described with the aid of a single time Hilbert space
$\mathfrak{H}_s$. [For
simplicity we consider here and in the sequel only quantum
systems without superselection rules.]
The observables associated with the system are
identified with self adjoint operators on the single time Hilbert
space $\mathfrak{H}_s$ and the quantum mechanical states of the
physical system with density operators on the single time
Hilbert space $\mathfrak{H}_s$. As is obvious from these
statements all observables and states are associated with a fixed time
(or slightly more general with a fixed spacelike hypersurface) and
there is no notion of observable associated with an extended
region of space time in standard quantum mechanics.
The time evolution is governed by certain
unitary operators on the single time Hilbert space
$\mathfrak{H}_s$. \\
The aim of the history approach (at least for the purpose of the
present investigation) is to formulate an intrinsically quantum
mechanical formalism in which observables and states are
associated with extended space time regions and in which time
plays a potentially subsidiary role. There have been a few
attempts in the literature
to extrapolate the usual Hilbert space formalism also to
situations involving observables associated with extended space
time regions in an \emph{ad hoc} way. However, it is not \emph{a priori}
clear whether such a simple strategy can be justified.
In contrast in the history approach (particularly in the approach
pioneered by Isham \cite{Isham94}) one proceeds along a different
route. Methodically what one is trying to do is to find a quantum
mechanical formalism involving space time observables and states
by starting with suitably reformulating standard quantum
mechanics. In the present paper we review the progress of this program
which has been made in the recent years \cite{Isham94} -
\cite{Schreckenberg96a}, \cite{Wright95} - \cite{Wright99}.

This article is structured as follows. In the first part (sections
2 and 3)
we review the general framework and perspective of so-called
history quantum theories. Specifically we shall discuss the
history reformulation of non-relativistic quantum mechanics in the
case that the underlying single time Hilbert space $\mathfrak{H}_s$
is finite dimensional and introduce the notion of decoherence
functional which represent the states in the present approach.
This history reformulation of standard quantum mechanics serves as
motivation for Isham's algebraic axiomatization of general history
quantum theories. The
temporal reformulation of standard quantum mechanics
suggests to consider general history quantum theories for which
the space of histories is given
by (i) the lattice of projection operators on some Hilbert space or --
slightly more general -- (ii) the set of projection operators in some
von Neumann algebra. The second main part of this article (section 4) is
devoted to the representation theory of decoherence functionals both in
general history theories and in the history reformulation of
standard quantum mechanics. If the underlying history Hilbert
space $\mathcal{H}$
is finite dimensional, a complete
classification of decoherence functionals has been given by Isham,
Linden and Schreckenberg \cite{IshamLS94}
with their so-called ILS-theorem which
establishes a one-to-one correspondence between bounded
decoherence functionals and certain trace class operators.
In the case that the history Hilbert space $\mathcal{H}$ is
infinite dimensional, we shall be concerned with the problem for
which decoherence functionals the ILS-theorem can be generalized.
Moreover, we shall discuss the natural representations of bounded
decoherence functionals as sesquilinear forms with a natural
representation on a Hilbert space.
Since the history reformulation of standard non-relativistic
quantum mechanics is the motivating example for the history
approach we shall be particularly interested in representations of
the standard decoherence functional $d_\rho$ associated
with the initial state $\rho$ in
standard quantum mechanics.
We shall conclude this paper with a discussion of our results for
the general framework for history theories proposed by Isham
and put forward a modified axiomatization of
history quantum theories.

\subsubsection*{Notations and conventions}
Throughout this work we will make use of Dirac's well-known ket
and bra notation to denote vectors in Hilbert space and dual
vectors in the dual Hilbert space respectively. We adopt the
convention that inner products of Hilbert spaces are linear in the
second variable and conjugate linear in the first variable.

Throughout this work $\frak H$ and $\mathcal{H}$ denote Hilbert spaces,
${\cal
P}(\HH)$ denotes the lattice of all projection operators on a Hilbert space
$\frak H$, ${\cal B}({\frak H})$ denotes the set
of all bounded operators on $\frak H$ and $\mathcal{K}(\mathfrak{H})$
denotes the set of compact operators on $\mathfrak{H}$.
The tensor product of the two Hilbert spaces $\mathcal{H}_1$ and
$\mathcal{H}_2$ is denoted by $\mathcal{H}_1 \otimes
\mathcal{H}_2$. The algebraic tensor product of $\mathcal{K}(\mathcal{H}_1)$
with $\mathcal{K}(\mathcal{H}_2)$ is denoted by $\mathcal{K}(\mathcal{H}_1)
\otimes_{alg} \mathcal{K}(\mathcal{H}_2)$.
The symbol $\mathfrak{H}_s$
denotes always the single time Hilbert space in standard quantum
mechanics.

\section{The history reformulation of standard quantum mechanics}
\subsection{Homogeneous histories}

Our starting point towards a formal definition of the notion of
history is the observation that -- by virtue of the spectral theorem --
every observable in standard quantum mechanics can be
disintegrated in two-valued yes-no observables which are
represented by projection operators on the single time Hilbert
space $\mathfrak{H}_s$. In a first step towards the history
reformulation of standard quantum mechanics one considers finite
sequences of projection operators \cite{Griffiths84}
\[ h = P_{t_1}, P_{t_2}, \cdots,
P_{t_n} \] labeled by a discrete set of time parameters $\{ t_1,
\cdots, t_n \}$. We call such a sequence $h$ a \emph{homogeneous
history}.
Operationally one may think of such a sequence as representing a
sequence of possible measurement outcomes.

Standard quantum mechanics suggests the following Ansatz for the
quantum mechanical probability of a homogeneous history $h$ in the
state $\rho$ which
we denote by the symbol $d_\rho(h,h)$
\begin{equation}
d_\rho(h,h) = \mathrm{tr}_{\mathfrak{H}_s}(P_{t_n} \cdots P_{t_2}
P_{t_1} \rho P_{t_1} P_{t_2} \cdots P_{t_n}).
\end{equation}
This expression coincides with the formula for the probability of
the sequence of measurement outcomes corresponding to the sequence of
projections $\{ P_{t_1}, \cdots, P_{t_n} \}$ in a measurement situation
\cite{Wigner63}.
[Notice
that we are working in the Heisenberg picture here and suppress
for notational simplicity the unitary time evolution operators in
the expression for the probability.]

At this stage we are facing a list of problems
\begin{enumerate}
\item The space of all homogeneous histories carries no obvious
``nice'' and simple mathematical structure and, particularly,
in general it is
not obvious what the appropriate mathematical representatives
corresponding
to propositions like ``\emph{the history $h$ or the history $k$ is
realized},'' ``\emph{the histories $h$ and $k$ are both realized}''
and ``\emph{the history $h$ is not realized}'' are.
\item There is no notion of ``sum'' of homogeneous histories and
\item therefore there is no additive probability measure on the
space of homogeneous histories.
\end{enumerate}

\subsection{Temporal quantum logic}
The first two problems have been solved by Isham in
\cite{Isham94}. He observed that every homogeneous history $h =
\{ h_{t_i} \}$ can be canonically mapped to some projection
operator on the $n$-fold tensor product Hilbert space $\otimes_{t_i}
\mathfrak{H}_{t_i}$ (where $\mathfrak{H}_{t_i} = \mathfrak{H}_s$
for all $i$) of the single time Hilbert space $\mathfrak{H}_s$ by
itself via \[ h \simeq \{ h_{t_i} \} \mapsto h_{t_1} \otimes \cdots
\otimes  h_{t_n}. \]
Now we observe that the space $\mathcal{P}(\otimes_{t_i}
\mathfrak{H}_{t_i})$ of all projections on this tensor
product Hilbert space carries the structure of a lattice.
The central postulate in Isham's temporal quantum logic is to
identify all projections in $\mathcal{P}(\otimes_{t_i}
\mathfrak{H}_{t_i})$ with physical histories.
The lattice theoretical operations in $\mathcal{P}(\otimes_{t_i}
\mathfrak{H}_{t_i})$ then provide a natural solution to the first
two problems mentioned above.

The space of all histories can then be identified with the
following direct limit
\[ \mathcal{P} := \lim \{ \mathcal{P}(\otimes_{t_i \in I}
\mathfrak{H}_{t_i}) \vert I \subset \mathbb{R} \mbox{ finite } \}
\] All histories in $\mathcal{P}$
which are not homogeneous are also called
\emph{inhomogeneous histories}.

Now after we have identified the space of histories in the
history reformulation of standard quantum mechanics, we are
interested in what the dual notion representing the states is.

\subsection{Decoherence functionals}

A \emph{decoherence functional} $d$ is a bivariate, complex valued
functional $d :
\mathcal{P} \times \mathcal{P} \to \mathbb{C}$
such that for all $\alpha, \alpha',\beta \in
\mathcal{P}$ with $\alpha \perp \alpha'$
\begin{itemize} \item $d(\alpha,\alpha) \in
\mathbb{R}$ and
$d(\alpha,\alpha)
\geq 0$. \item $d(\alpha,\beta) =
d(\beta,\alpha)^*$. \item
$d(1,1) =1$ and $d(0,\alpha)
=0$, for all
$\alpha$. \item $d(\alpha
\vee \alpha', \beta) =
d(\alpha,\beta) + d(\alpha',\beta)$.
\end{itemize}

The idea behind the positivity requirement for the diagonal values
of $d$ is that $d(\alpha, \alpha)$ represents the probability of
the history $\alpha$.

The prime and motivating example for a decoherence functional with
the above list of properties is the decoherence functional $d_\rho$
in standard quantum mechanics associated with the initial state
$\rho$ which is defined for homogeneous histories $h \simeq \{
h_{t_i} \}$ and $k \simeq \{ k_{t_j} \}$ by
\begin{equation} \label{drho}
d_\rho(h,k) := \mathrm{tr}(h_{t_n}
h_{t_{n-1}} \cdot \cdot
\cdot h_{t_1} \rho k_{t_1} \cdot \cdot \cdot k_{t_n}). \end{equation}
This is a modest generalization of the above expression for the
probability $d_\rho(h,h)$ of some history $h$.
When the single time Hilbert space $\mathfrak{H}_s$ is finite dimensional,
then the such defined decoherence functional $d_\rho$
can uniquely be extended to a bi-additive function on the set of
all histories $\mathcal{P}$ as will be shown below.
However, we shall also argue below that $d_\rho$ cannot be extended to
a finitely valued functional on the set $\mathcal{P}$
of all histories if the
single time Hilbert space is infinite dimensional.

\subsection{ILS-representation for $d_\rho$}
The decoherence functional $d_\rho$ associated with the initial
state $\rho$
in standard quantum mechanics defined for homogeneous histories by
Equation (\ref{drho}) admits a so-called
Isham-Linden-Schreckenberg representation (or more shortly
ILS-representation) \cite{IshamLS94}.
This means that for all $n$-time histories $h =
h_{t_1} \otimes \cdots \otimes h_{t_n}$ and $k = k_{t_1} \otimes \cdots
\otimes k_{t_n}$ there exists a trace class operator $\mathfrak{X}_\rho$
on the $2n$-fold tensor product $\mathcal{H}_n \otimes \mathcal{H}_n =
\mathfrak{H}_{t_1} \otimes \cdots
\otimes \mathfrak{H}_{t_n} \otimes \mathfrak{H}_{t_1} \otimes
\cdots \otimes \mathfrak{H}_{t_n}$ of the single time Hilbert
space $\mathfrak{H}_s$ by itself such that $d_\rho$ can be
represented as
\begin{equation}
d_\rho(h,k) = \mathrm{tr}_{\mathcal{H}_n \otimes \mathcal{H}_n} \left(h
\otimes k \mathfrak{X}_\rho \right). \label{drho2}
\end{equation}
The operator $\mathfrak{X}_\rho$ depends both on the initial state
$\rho$ and on the unitary time evolution operator $U(t,t')$. The
dependence on $U$ can explicitly split up

\[
\mathfrak{X}_\rho = \left(U^{\dagger}_{t_1,t_2, ..., t_n} \otimes
U^{\dagger}_{t_1,t_2, ..., t_n} \right) {\frak Y}_{\varrho}
\left( U_{t_1,t_2,
..., t_n} \otimes U_{t_1,t_2, ..., t_n}) \right) \]
where
\[ U_{t_1,t_2, ..., t_n} := U(t_0,t_1) \otimes U(t_0, t_2) \otimes
\cdot \cdot \otimes U(t_0, t_n). \]
We are left with an operator $\mathfrak{Y}_\rho$ depending only on
the initial state $\rho$. The operator $\mathfrak{Y}_\rho$ admits
a representation as a series
\begin{eqnarray*} {\frak Y}_{\rho} = \sum_{i_1, ..., i_{2n}} &
\omega_{i_{1}} & \left\{ \vert e^1_{i_1} \rangle \langle
e^{2n}_{i_{2n}} \vert \otimes \vert e^{2n}_{i_{2n}} \rangle
\langle e^{2n-1}_{i_{2n-1}} \vert \otimes \cdot \cdot \cdot \otimes
\vert e^{n+2}_{i_{n+2}} \rangle \langle e^{n+1}_{i_{n+1}} \vert \otimes
\right. \\
& & \hspace*{0.2cm} \left. \otimes \vert e^{2}_{i_{2}} \rangle
\langle e^{1}_{i_{1}} \vert \otimes \vert e^{3}_{i_{3}}
\rangle \langle e^{2}_{i_{2}} \vert \otimes \cdot \cdot \cdot
\otimes
\vert e^{n+1}_{i_{n+1}} \rangle \langle e^{n}_{i_{n}}
\vert \right\}. \end{eqnarray*}
The $\omega_i$ are determined by the spectral resolution of $\rho
= \sum_i \omega_i \vert e^\rho_i \rangle \langle e^\rho_i \vert$.
The orthonormal bases $\{
\vert e^j_{i_j} \rangle \}$, $j \in \{2, ..., 2n \}$ are
completely arbitrary, whereas $\vert e^{1}_{i}
\rangle = \vert e^{\rho}_i \rangle $ for all $i$.

If we restrict ourselves to homogeneous histories $h$ and $k$ then
the ILS-representation in Equation (\ref{drho2}) is valid both when
the single time Hilbert space is finite or infinite dimensional.
In the finite dimensional case $\mathfrak{X}_\rho$ is a trace class operator.
Thus, trivially, $d_\rho$ can be extended to the set $\mathcal{P}$
of all histories.
In the infinite dimensional case it can be shown
\cite{RudolphW98b}  that $\mathfrak{X}_\rho$ is only a bounded
operator. Therefore in the infinite dimensional case $d_\rho$ can
in general not be extended to the space of all histories. We shall
come back to this issue below.

\subsection{Consistent sets of histories}
It remains to solve the third problem mentioned in section 2.1
that there is no additive probability measure on the space of
homogeneous histories. We have seen that in the history approach
the states are identified with decoherence functionals. Again,
these decoherence functionals do not define a probability measure
on the set of all histories.

The situation is analogous to standard single time quantum
mechanics. Here the states are given by density operators on the single
time Hilbert space which do
not induce probability measures on the set of \emph{all} observables.
In standard quantum mechanics we call a set of observables
\emph{compatible} if the state induces a joint probability measure
on the set of possible values of the observables. It is well known
that a set of observables is compatible if and only if the
associated self adjoint operators are pairwise commuting
(particularly, the notion of compatibility of observables is
independent of the state).

Generalizing this point of view to the histories approach one
calls a set of histories \emph{consistent} if the decoherence
functional induces a probability measure on this set of histories.
It is easy to prove that a Boolean sublattice $\mathcal{C}$
of $\mathcal{P}$ is
consistent if and only if $\mathrm{Re }\, \, d_\rho(h,k) = 0$ for all
orthogonal $h,k \in \mathcal{C}$.
The consistent sets of histories are thus the generalizations of
commuting, compatible observables in standard quantum mechanics and
the existence of several mutually inconsistent consistent sets is
just the expression of the complementarity principle in the
histories approach.

\section{General history theories}
The history reformulation of standard quantum mechanics in finite dimensions
reviewed above has led Isham \cite{Isham94} to his axiomatic framework for
general history quantum theories. According to his framework a general
history theory is
characterized by two sets.

First there is the space of histories $\mathcal{U}$ which carries
the structure of a lattice, an orthoalgebra, a D-poset or another
algebraic structure
such that (i) there is a partial order defined, (ii)
there is a least element $0$
and a greatest element $1$ with respect to this partial order
and such that (iii) there is a notion of orthogonality
between elements (denoted by $\perp$) and
a notion of sum (denoted by $\oplus$)
for orthogonal elements.
The tentative physical interpretation of the histories
is that they represent propositions about events in extended
regions of space time.

Dual to the space of histories is the space of decoherence
functionals which represent the generalized states in the history
approach. Recall that decoherence functionals are bivariate,
complex valued functionals $d: \mathcal{U} \times \mathcal{U} \to
\mathbb{C}$ such that for all $\alpha, \alpha',\beta \in
\mathcal{U}$ with $\alpha \perp \alpha'$
\begin{itemize} \item $d(\alpha,\alpha) \in
\mathbb{R}$ and
$d(\alpha,\alpha)
\geq 0$. \item $d(\alpha,\beta) =
d(\beta,\alpha)^*$. \item
$d(1,1) =1$ and $d(0,\alpha)
=0$, for all
$\alpha$. \item $d(\alpha
\oplus \alpha', \beta) =
d(\alpha,\beta) + d(\alpha',\beta)$.
\end{itemize}

The history reformulation of standard quantum mechanics in finite
dimensions suggests that a natural choice for the space of
histories is given by the set of projection operators on some
(history) Hilbert space $\mathcal{H}$
or von Neumann algebra $\mathcal{A}$. In the rest of this paper we
shall exclusively consider these two cases.
The second part of this paper is devoted to the problem what can be
said about the structure of the space of decoherence functionals
for these choices of the space of histories. \newpage

\section{The representation theory of decoherence functionals}

\subsection{Finite dimensional history Hilbert spaces}
First let us consider an abstract history theory (as described in
section 2.5) for which the space of histories is given by set of
projection operators $\mathcal{P}(\mathcal{H})$
on some finite dimensional Hilbert space $\mathcal{H}$
(with dimension greater than two).

In this case the classification problem for decoherence
functionals has been completely solved by Isham, Linden and
Schreckenberg \cite{IshamLS94}. According to their result there
exists a one-to-one correspondence between uniformly bi-continuous
decoherence functionals
$d$ for $\mathcal{H}$
and trace class operators $\mathfrak{X}$
on $\mathcal{H} \otimes \mathcal{H}$
according to the rule
\begin{equation} d(p,q) =
\mathrm{tr}_{\mathcal{H} \otimes
\mathcal{H}} \left((p \otimes q) \mathfrak{X}
\right), \label{E13} \end{equation}
for all projections $p,q \in
\mathcal{P}(\mathcal{H})$
with the restriction that
\begin{itemize} \item
$\mathrm{tr}_{\mathcal{H} \otimes
\mathcal{H}} \left((p \otimes q)
\mathfrak{X} \right) =
\mathrm{tr}_{\mathcal{H} \otimes
\mathcal{H}} \left((q \otimes p)
\mathfrak{X}^* \right)$;
\item $\mathrm{tr}_{\mathcal{H}
\otimes \mathcal{H}}((p
\otimes p) \mathfrak{X})
\geq 0$;
\item $\mathrm{tr}_{\mathcal{H} \otimes
\mathcal{H}}(\mathfrak{X})
= 1$. \end{itemize}
In particular every such decoherence functional is bounded. This
is result is often also referred to as the
\emph{Isham-Linden-Schreckenberg theorem} (or more shortly the
ILS-theorem).

\subsection{Infinite dimensional history Hilbert spaces}
\subsubsection{ILS-type representations}
A question which arises immediately is what can be said for
history theories where the space of histories is given by the set
of projections on some infinite dimensional Hilbert space
$\mathcal{H}$.

In the sequel we will make use of the
following theorem which is a
special case of a more general result
proved in Wright \cite{Wright95}.
\begin{theo} \label{blabla} Let $\mathcal{H}$
be a Hilbert space which is either
infinite
dimensional or of finite dimension
greater than two.
Then a decoherence functional
$d$ on $\mathcal{H}$ can be extended (uniquely) to a
bounded bilinear form
$\mathcal{D} : \mathcal{B}(\mathcal{H})
\times
\mathcal{B}(\mathcal{H}) \to \mathbb{C}$ if,
and only if, $d$ is bounded. \label{T1} \end{theo}

An immediate consequence of Theorem \ref{T1} is then that,
by the fundamental property
of the algebraic tensor product, there
is a unique linear functional
$\beta : \mathcal{K}(\mathcal{H})
\otimes_{alg} \mathcal{K}(\mathcal{H})
\to \mathbb{C}$ on the algebraic tensor product of
$\mathcal{K}(\mathcal{H})$ by itself such that
\begin{equation} \beta(x \otimes y) =
\mathcal{D}(x,y), \label{E2}
\end{equation} \nopagebreak
for all $x, y \in
\mathcal{K}(\mathcal{H})$.
In particular $d(p, q) =
\beta(p \otimes q)$ for all projections $p$
and $q$ in
$\mathcal{K}(\mathcal{H})$.

The functional $\beta$ can now be used to completely characterize
the set of decoherence functionals in the infinite dimensional
case admitting an ILS-representation.

\paragraph{Tensor bounded decoherence functionals \\ \\}
\hspace*{-2.711ex} \noindent\textbf{Definition}
\emph{The decoherence functional
$d$ is said to be \textsf{tensor
bounded} if the associated functional
$\beta$ is bounded on
$\mathcal{K}(\mathcal{H}) \otimes_{alg}
\mathcal{K}(\mathcal{H})$,
when $\mathcal{K}(\mathcal{H})
\otimes_{alg} \mathcal{K}(\mathcal{H})$
is equipped with its unique
pre-$C^*$-norm induced by the operator norm on $\mathcal{B}(\mathcal{H}
\otimes \mathcal{H})$.}

\begin{theo} \label{T2}
Let $\mathcal{H}$ be a Hilbert space
which is not of dimension
two. Let $d$ be a bounded decoherence
functional for $\mathcal{H}$.
Then
$d$ is tensor bounded if, and only if,
there exists a trace class
operator $\mathfrak{X}$ on
$\mathcal{H} \otimes \mathcal{H}$ such that
\begin{equation} d(p,q) =
\mathrm{tr}_{\mathcal{H} \otimes
\mathcal{H}} ((p \otimes q)
\mathfrak{X}) \label{E3} \end{equation}
for all projections $p$ and $q$
in $\mathcal{P}(\mathcal{H})$.
\end{theo}
Thus we conclude that in the infinite dimensional case a bounded decoherence
functional $d$
admits an ILS-representation if and only if it is tensor bounded.

\paragraph{Tracially bounded decoherence functionals \\ \\}
There is another physically important class of decoherence
functionals, the so called tracially bounded decoherence
functionals. \\

\noindent\textbf{Definition}
\enlargethispage{0.7cm}
\emph{A decoherence functional
$d$ is said to be
\textsf{tracially bounded}
if it is bounded and, when $\beta$
is the corresponding linear functional
on $\mathcal{K}(\mathcal{H}) \otimes_{alg}
\mathcal{K}(\mathcal{H})$,
there exists a constant $C$ such that,
for each unit
vector $\xi$ in $\mathcal{H} \otimes_{alg}
\mathcal{H}$,
$\vert \beta \left( \vert \xi \rangle \langle \xi \vert \right)
\vert \leq C$.} \\

Then we have the following theorem
\begin{theo} \label{bla} Let the decoherence
functional $d$ be tracially
bounded for
$\mathcal{H}$ where
$\mathcal{H}$ is separable and of
dimension greater than two. Then there exists a unique
bounded linear operator
$\mathfrak{M}$ on $\mathcal{H}
\otimes \mathcal{H}$ such that
\begin{equation} d(p,q) =
\mathrm{tr}_{\mathcal{H} \otimes
\mathcal{H}} \left( \mathfrak{M}(p
\otimes q) \right) \label{E8}
\end{equation}
whenever $p$ and $q$ are finite rank
projections on $\mathcal{H}$.
Let $d$ be moreover countably
additive, then
whenever $p$ and $q$
are projections
in $\mathcal{P}(\mathcal{H})$ and
$\{p_n \}_{n \in \mathbb{N}}$
and $\{q_n \}_{n \in \mathbb{N}}$
are, respectively,
orthogonal families of finite rank
projections with
$p = \sum_{n \in \mathbb{N}} p_n$ and
$q = \sum_{n \in \mathbb{N}} q_n$,
\begin{equation} d(p,q) =
\sum\limits_{i=1}^{\infty}
\sum\limits_{j=1}^{\infty}
\mathrm{tr}_{\mathcal{H}
\otimes \mathcal{H}} \left((p_i
\otimes q_j) \mathfrak{M} \right).
\label{E0}
\end{equation} \end{theo}
Thus we see that tracially bounded decoherence functionals admit a
pseudo-ILS-representation by some bounded operator as in Equation
(\ref{E8}).

\subsubsection{The standard decoherence functional $d_\rho$ in
infinite dimensions}

Tracially bounded decoherence functionals are of particular
interest since the standard decoherence functional $d_\rho$ in the
history reformulation of standard
quantum mechanics is tracially bounded.

The proof is quite easy \cite{RudolphW98b}.
In finite dimensions this is trivial. In
infinite dimensions the argument is as follows. For
simplicity of notation we
consider only two time histories, the general case is analogous.
First recall the ILS-representation of $d_\rho$ valid for pairs of
homogeneous histories $p,q$
 \[ d_\rho(p,q) =
\sum_{j_1, \cdots, j_{4}=1}^{\dim {\frak H}_s}
\omega_{j_1}
\left\langle e^{4}_{j_{4}} \otimes e^3_{j_3}
\otimes \psi_{j_1} \otimes e^2_{j_2},
(p \otimes q) (\psi_{j_1} \otimes e^{4}_{j_{4}} \otimes
e^2_{j_2} \otimes e^3_{j_3})
\right\rangle.  \]  \normalsize
It is easy to see that the series
still converges if we replace $p \otimes q$
by a compact operator of rank one. This implies that we can
define a sesquilinear form $S_\rho$ by
\[ S_\rho(\xi, \eta) =
\sum_{j_1, \cdots, j_{4}=1}^{\dim {\frak H}_s}
\omega_{j_1}
\left\langle e^{4}_{j_{4}} \otimes e^3_{j_3}
\otimes \psi_{j_1} \otimes e^2_{j_2}, \xi \right\rangle
\left\langle \eta, \psi_{j_1} \otimes e^{4}_{j_{4}} \otimes
e^2_{j_2} \otimes e^3_{j_3} \right\rangle. \] \normalsize
The Cauchy-Schwarz inequality implies that $S_\rho$ is bounded:
$ \vert S_\rho(\xi, \eta) \vert \leq \Vert \xi \Vert \Vert \eta
\Vert$ which in turn implies
$\vert \beta_\rho(p_\xi) \vert = \vert S_\rho(\xi, \xi)
\vert \leq 1$. This implies that there exists a bounded operator
$\mathfrak{X}_\rho$ such that $S_\rho(\xi, \eta) = \langle \eta,
\mathfrak{X}_\rho \xi \rangle$ and by straightforward computation
one verifies that $\mathfrak{X}_\rho$ coincides with the
ILS-operator associated with $d_\rho$ in section 2.4. Thus $d_\rho$
admits a pseudo-ILS-representation as in Theorem \ref{bla} with
$\mathfrak{M}$ replaced by $\mathfrak{X}_\rho$ and
since $d_\rho$ is moreover countably additive the analogue of
Equation \ref{E0} is also satisfied (whenever well-defined).

As we shall see below $d_\rho$ is in general not bounded (and not
even finitely valued) on the space of all histories $\mathcal{P}$.
Thus we cannot apply Theorem \ref{blabla} to infer the existence of the
functional $\beta_\rho$ associated with $d_\rho$. However, from
the ILS-series for $d_\rho$ we can directly infer the existence of
$\beta_\rho$ on a suitably smaller chosen domain of definition.

\paragraph{Non-existence of a finitely valued extension of $d_\rho$
\\ \\}

As already mentioned repeatedly if
the single time Hilbert space is infinite dimensional, then the
standard decoherence functional $d_\rho$ defined on homogeneous
histories by Equation (\ref{drho}) cannot be extended to a
finitely valued functional on the set of all projection operators
on the tensor product Hilbert space.
We assume for simplicity that the single time
Hilbert space is separable. \\
Consider the ILS-representation for $d_\rho$
in Equation (\ref{drho2}). For
simplicity of notation we consider the case $n=2$. We define
\begin{equation} \label{decf3} D_\rho(p,q) =
\sum_{j_1, \cdots, j_{4}=1}^{\dim {\frak H}}
\omega_{j_1}
\left\langle e^{4}_{j_{4}} \otimes e^3_{j_3}
\otimes \psi_{j_1} \otimes e^2_{j_2},
(p \otimes q) (\psi_{j_1} \otimes e^{4}_{j_{4}} \otimes
e^2_{j_2} \otimes e^3_{j_3})
\right\rangle,  \end{equation} for all
histories $p, q \in {\cal P}({\frak H}_{t_1} \otimes
{\frak H}_{t_2})$ for which the sum
converges. Now choose $e^4_j = e^3_j = e^2_j = \psi_j$ for all $j$. Fix
$i_1$ and let $\varphi_i := \frac{1}{\sqrt{2}} \left(
\vert \psi_i \otimes \psi_{i_1} \rangle + \vert \psi_{i_1} \otimes \psi_i
\rangle \right)$
for every $i \in {\Bbb N} \backslash \{ i_1 \}$.
Then clearly $\varphi_i \perp \varphi_j$ if $i \neq j$.
Set $f_{j_1,j_2,j_3}(q) := \langle \psi_{j_1} \otimes \psi_{j_2},
q (\psi_{j_2} \otimes \psi_{j_3}) \rangle$, then an easy
computation shows that \[ D_\rho(P_{\varphi_i}, q) = \frac{1}{2}
\sum_{j_2} \left( \omega_{i_1} f_{i_1,j_2,i_1}(q) + \omega_i
f_{i,j_2,i}(q) \right), \] for $i \neq i_1$ where $P_{\varphi_i}$
denotes the
projection operator onto the subspace spanned by $\varphi_i$. Put
$P=\sum_{i \neq i_1} P_{\varphi_i}$, then clearly the expression in
Equation (\ref{decf3}) for $D_\rho(P,q)$ does not
converge for arbitrary $q$.

This proves that if the single time Hilbert space is infinite
dimensional, there does not exist a finitely valued extension of $d_\rho$
to the set of histories $\mathcal{P}$.

\subsubsection{Bounded decoherence functionals}
We now return to our discussion of representations of decoherence
functionals in general history theories. We have identified above
the classes of decoherence functionals admitting an
ILS-representation and a pseudo-ILS-representation respectively.
It is also of some interest what can be said about general bounded
decoherence functionals. Although a bounded decoherence functional
in general does not admit an ILS-representation
they can be approximated by a series of ILS-representable
decoherence functionals in the following sense.

\begin{prop} Let $\mathcal{H}$ be
a Hilbert space with $\dim(\mathcal{H}) > 2$ and let
\mbox{$d : \mathcal{P}(\mathcal{H})
\times \mathcal{P}(\mathcal{H}) \to
\mathbb{C}$} be a bounded decoherence
functional for $\mathcal{H}$.
Then there exist families of trace
class operators
$\{ \mathfrak{X}_i \}_{i \in I}$ and
$\{ \mathfrak{Y}_i \}_{i \in I}$
on $\mathcal{H}$, where, for each
$x$ and $y$ in
$\mathcal{K}(\mathcal{H})$,
$\sum_{i \in I} \vert
\mathrm{tr}_{\mathcal{H}}(x
\mathfrak{X}_i) \vert^2$
and $\sum_{i \in I} \vert
\mathrm{tr}_{\mathcal{H}}(y
\mathfrak{Y}_i) \vert^2$ are convergent
and, for all $p,q \in
\mathcal{K}(\mathcal{H})$,
\begin{equation}
d(p,q) = \sum_{i \in I}
\mathrm{tr}_{\mathcal{H} \otimes \mathcal{H}}
\left(p \otimes q \left(\mathfrak{X}_i \otimes
\mathfrak{X}^{*}_i -
\mathfrak{Y}_i \otimes
\mathfrak{Y}^{*}_i \right) \right),
\end{equation}
where the infinite series is absolutely
convergent.
\label{P2} \end{prop}

\subsection{Representations as sesquilinear forms}

\paragraph{Bounded decoherence functionals \\ \\}
There is an alternative representation theorem for bounded
decoherence functionals in general history theories as bounded
sesquilinear forms on a Hilbert space due to Wright
\cite{Wright95}. This is valid also for history theories over
von Neumann algebras (with no type $I_2$ direct summand).

\begin{theo}
Let {$A$} be a von Neumann algebra with no type $I_2$ direct
summand and {$d: \mathcal{P}(A) \times \mathcal{P}(A) \to \mathbb{C}$}
a bounded decoherence functional. Then there exists a map {$x \mapsto
[x]$} from {$A$} into a dense subspace of a Hilbert space
{$\mathcal{H}$} and a self adjoint operator {$T$} on
{$\mathcal{H}$} such that { \[ D(x,y) = \langle T [x], [y]
\rangle \]}
is an extension of {$d$}. \\
Alternatively there exist semi inner products {$\langle \cdot,
\cdot \rangle_1$} and {$\langle \cdot,
\cdot \rangle_2$} on {$\mathcal{H}$}
such that { \[ d(p,q) = \langle p, q
\rangle_1 - \langle p, q \rangle_2 \]} \end{theo}
The proof makes use of the profound Haagerup-Pisier-Grothendieck
inequality to associate a state (in the $C^*$ algebraic sense)
with the decoherence functional. The Hilbert space is then
constructed via a GNS-type construction.

\paragraph{Standard decoherence functional \\ \\}

As shown above in standard quantum mechanics
the standard decoherence functional $d_\rho$ does
not admit a finitely valued extension to the set of all histories
in Isham's framework. Thus there is also no hope to represent it
as a bounded sesquilinear form on some Hilbert space. However,
there is a natural representation of the standard decoherence
functional as an unbounded sesquilinear form which in brief can be
constructed as follows.

If {$\mathfrak{H}_s$} is infinite dimensional, then
{$d_\rho$} can be extended to bilinear functional on
{$\mathcal{B}(\mathfrak{H}_s) \otimes_{alg} \cdots \otimes_{alg}
\mathcal{B}(\mathfrak{H}_s)$} ($n$ times) as
{ \[ D_\rho (b, b') := \mathrm{tr}(\Pi(b')^\dagger
\Pi(b) \rho) \]} where $\Pi$ is defined on homogeneous elements by
{$\Pi(b_1
\otimes \cdots \otimes b_n) = b_n \cdots b_1$} and extended to all of
$\mathcal{B}(\mathfrak{H}_s) \otimes_{alg} \cdots \otimes_{alg}
\mathcal{B}(\mathfrak{H}_s)$ by linearity.
\begin{theo}
There exists a
Hilbert space {$\mathcal{H}$} and a linear operator
{$R_\rho$} from $\mathcal{B}(\mathfrak{H}_s) \otimes_{alg}
\cdots \otimes_{alg}
\mathcal{B}(\mathfrak{H}_s)$ into a dense subspace of {$\mathcal{H}$}
such that { \[ D_\rho (b, b') = \langle R_\rho(b'), R_\rho(b)
\rangle \]} for all {$b,b' \in
\mathcal{B}(\mathfrak{H}_s) \otimes_{alg} \cdots \otimes_{alg}
\mathcal{B}(\mathfrak{H}_s)$}. (Here $\langle \cdot, \cdot \rangle$ denotes
the inner product in $\mathcal{H}$.) \\
{$R_\rho$} is unbounded if and only if {$\mathfrak{H}_s$} is infinite
dimensional. \end{theo}

\section{General history theories II}

At this stage it is worthwhile to recall that Isham's axiomatic
framework for general history theories sketched in section 3
was motivated by the
history reformulation of standard quantum mechanics in
\emph{finite} dimensions.

The results reported in this paper, in particular the
negative result that the standard decoherence functional in
infinite dimensions cannot be extended to the space of ``all''
histories in Isham's framework on the one hand and the positive
result that the standard decoherence functional admits
nevertheless a natural
representation as an unbounded sesquilinear form on some Hilbert
space on the other hand, indicate that Isham's
axiomatic framework needs to be modified.

We shall conclude this paper by indicating the in our opinion
appropriate structure.

According to our proposal a
general history theory is characterized by two sets.
\begin{itemize}
\item Firstly, the set of propositions which is embedded into a Hilbert
space $\mathcal{H}$.
The propositions are interpreted in physical terms as
propositions about events in extended regions of space-time.

\item Secondly, the set of states which are identified with bounded or
unbounded sesquilinear forms $s$ on the Hilbert space $\mathcal{H}$.
\end{itemize}

The probability of a proposition $x \in \mathcal{D}(s)$ in the
domain of definition of some sesquilinear form is given by
$s(x,x)$.
This framework for temporal quantum theories has been discussed in
more detail in \cite{Rudolph98a}.

\subsection*{Acknowledgements}
Oliver Rudolph is a Marie Curie Research Fellow and carries
out his research at Imperial College as part of a European
Union training project financed by the European Commission
under the
TMR programme. \\
I am very grateful to Professor C.J.~Isham for his support of my
work. I also should like to acknowledge my indebtedness to my
collaborator, Professor J.D.M.~Wright. Without the privilege of
his cooperation and help I would have little to report here.

\end{document}